\documentstyle{article}

\newcommand{\ra}{\rangle}
\newcommand{\la}{\langle}
\newcommand{\be}{\begin{eqnarray}}
\newcommand{\ee}{\end{eqnarray}}
\newcommand{\ben}{\begin{eqnarray*}}
\newcommand{\een}{\end{eqnarray*}}

\begin{document}
\large
\begin{center}
{\bf SCHR\"ODINGER'S  CAT AND   THE   PROBLEM \\ OF TWO CULTURES}\\
{\em Ivan Vakarchuk\\
Ivan Franko National University of Lviv, Department for Theoretical Physics\\
           12 Drahomanov St., Lviv, 79005, Ukraine}
\end{center}

{\small
A mathematical model of the interaction mechanism for the intuitive-imaginative and
heuristic-logical thinking responsible for the rift in the intellectual activity into
two cultures has been suggested. The said model proceeds from the assumption that human
thinking is based on the principles of the many-channel quantum-mechanical logic  of the "both ... and"
type surpassing the rigid confines of the "either ... or" type classical logic. The aggregate product
of the person equally endowed in the said two-cultural space  has been  calculated. The interferential
part of the latter achieves its maximum at the extreme values of such parameters as the inter-state
exchange frequency and the difference of the states phases.
}

\vspace*{2cm}

{\em
\hspace*{4.0cm}If we want to gauge the sky, the earth and the seas,\\
\hspace*{4.2cm} we must first
of all measure ourselves.}\\
\hspace*{11.0cm} Hryhorij Skovoroda\\
\\
\hspace*{4.2cm}
{\it Philosophy is neither a science nor a perceptual sermon.}\\
\hspace*{11cm} Martin Heidegger

\vspace*{1.3cm}

\begin{center}
{\bf I}
\end{center}

Let us begin to talk about the problem of two cultures  with a survey of one of its
aspects which is quite apparent in the university setting. When
considering the phenomenon of two cultures we shall venture to suggest
one of the possible mechanisms regulating the interaction between the
heuristic-logical and intuitive-imaginative thinking [1].

Each of us is aware of being different from everybody else. We are trying to measure this extent of otherness all the
time intuitively. Each
scholar tends to give some thought to the fact that his creative work and
intellectual activity happen to be quite different from what, say, writers
or actors are busy with. This is a well-known fact which has always been
intuitively recognized. It was also quite aptly worded by an English writer and
physicist Charles  Snow  in the fifties of the last
century. He introduced the term {\it two cultures} by
which he meant the culture of those who in their own words deal with
creating and whose activity is concerned with imaginative thinking and
the culture which is created by those working  in the realm of exact
and  natural sciences.  Thus on the one pole we have the artistic
intelligentsia and on the other pole we have scientists.

We can formulate the definition of the two cultures in stricter
terms. All the people  around us fall under two groups,  namely those who  are
well-versed in maths and those who feel at home with something else. In everyday
life we are only too often confronted with a dichotomy that there are
people who like mathematics   and there are those who dislike it. By this
we do not mean to refer to abilities, a desire to work or exert effort on
oneself. Such endeavours seldom lead to the desired effect as a person
may be capable of creating something else falling back on the principles
of imaginative rather than rational thinking.

However, there are some who fall out of this dichotomy. We know
a number of people who are profusely endowed with the
capabilities
of either type. In what follows we shall speak about them.

A natural laboratory to study the mentioned phenomenon can be
found within the walls of the classical university. In the words of Karl
Jaspers the university as a phenomenon of culture brings together those
who are willing to cognize scientifically and to live spiritually. As can be
seen in this definition we can again trace  the two marked aspects  of
human activity just as well as in the quoted words of Martin Heidegger concerning philosophy.
A prime feature of the university environment lies in its encompassing
many cultures and many disciplines. Within  a classical university where
there is an atmosphere of interweaving  tones and halftones,
rich colours and subtle tinges of humanities with  exact as well as natural
sciences not a single teacher or scholar can stand fully
aloof from  the problem of two cultures.

\begin{center}
{\bf II}
\end{center}

    Sometimes people who are lucky to be endowed with the faculties
going in both directions, {\it i.e.},  along the path of intuitive creativity and
logical-intellectual thought cannot fully implement these capabilities
which may lead to dramatic or even tragic circumstances. In order not to
lose your talent you must have it, but far more importantly, you
must  have a chance to set it free. Developing talent and setting it free
takes some  purposeful steps. A gifted person is forced to make an effort
on one's own self in order to attain self-realization. A person that is gifted in
both these ways and that has a natural possibility to stay in the creative
as well as logical state can be very successful by crossing over between them
but at the same  time there is a risk that the said cross-overs might lead to
frustration. What factors are responsible for the specific outcome?

    We shall not make an attempt to discuss the formation of these
capabilities of a man at the level of physiology which might be motivated
by certain genes or sets of genes [3]. It is also quite probable  that
specific intellectual or creative mechanisms are inherited even though
their coming real is dependent upon a number of factors, including social
factors, active both immediately before and prospectively after the birth
of a human being.

    In order to get some indepth insight into the problem of two
cultures
I suggest that we concentrate on one of the possible ways of its
description which would be based on quantum mechanics principles. We
shall speak about a certain analogy enabling us to draw qualitative
conclusions and account for specific facts from real life. It may also go
beyond the scope of a mere analogy.

    The first hint at the analogy that the mechanisms of our thinking
are quantum mechanic in nature   comes from the fact that the retina of
the human eye, being a component of the human brain, is capable of
registering several   photons (light quanta) [3,4]. It appears that the
sensitivity of our eye is unexpectedly high.  Its physiological threshold
equals one photon which is sufficient for arousing the receptor. Yet for
the brain to be able to take in the message it is necessary to have
5$\div$8 photons. Thus the reception threshold
equals several photons even though there might be pre-threshold
reception. Our brain is set to motion by single light quanta. The
brain begins to work at the atomic level. For the latter the quantum
mechanics laws as well as the  quantum mechanics mechanism of the
phenomena are already fully binding even though the time, during which the
electron stays in the coherent  state, is substantially falling short of the
time necessary for conveying excitations between neurons.

We may adduce other examples to the effect that quantum
mechanics is binding on the microscopic scale that we are accustomed to.
They are the phenomena of super-fluidity and superconductivity or
else, even more clearly, that of the laser.

Falling back on what has been said so far it might well be possible
to suppose that the brain, being a macroscopic set-up, is likely to abide in
its functioning also by the laws of quantum mechanics. The latter
amounts to the many channel interferential  quantum mechanics principle
of the type of ``both  and"  surpassing the classical logic model of the
nature of ``either  or".

Should the Reader of these notes believe that the adduced
argumentation is insufficient or that  it is unacceptable, the author will
agree to the suggestion that this is just one of the possible microscopic level
modelings
of what is created by the observable phenomenon of two-culturalism.

\begin{center}
{\bf III}
\end{center}

Let us speak about those who are capable of combining these two
cultures. In view of what has been stated above let us try and model their
state as the state of the quantum system making use of the quantum
mechanics assumption of the cat of Schr\"odinger who is both alive and dead
at the same time.

Let us remember that in the wake of the creation of quantum
mechanics in the years 1925--1926 a set of paradoxes was suggested to
illustrate the seeming absurdity of its fundamental principles. The
paradox suggested  in 1935 by Erwin  Schr\"odinger who was one of the
founders of quantum mechanics says that the non-determined state of the
quantum particle, for instance a  photon,  penetrating the semi-silver
mirror with the probability of ``1/2" and reflecting from it with the same
``1/2" probability passes over the cat that is shut in a box. Should the
particle penetrate the mirror, it will switch on the device killing the cat.
Should it fail to penetrate the mirror the cat will stay alive. As the
particle which abides by the laws of quantum mechanics is in the super-positional
state, {\it i.e.}, in the ``both\ldots and" and not ``either\ldots  or" states,
the said cat also finds itself in the same superpositional
``alive-and-dead-at-the-same-time" state. Otherwise stated,  should the quantum particle
realize both the possibilities at the same time, the cat as a macroscopic
system must simultaneously  realize both its states at the same time. The
essence of the  paradox lies in the fact that after the photon gets into the
mirror the state of the cat  is seemingly non-determined. In true
fact, non-determinedness  disappears at the moment of the interaction of the
photon with the device.

The very idea about the possibility for a system to simultaneously
stay in two possible states at once is laid at the foundation of
the present study. The Schr\"odinger's cat paradox was referred to
with the purpose of illustrating a possible direct link between
micro- and macroworld. It shows that both the state of the cat and
our perception of two-culturalism are a consequence of the
processes taking place at the molecular level. The state of the
classical cat becomes determined at the moment of the interaction
of the photon with the device when the so-called reduction of its
state amplitude occurs. The state of two-culturalism which is
created by our central nervous system  is quantum by nature just
like quantum in principle are such phenomena as super-fluidity of
liquid helium or laser radiation.

Let us now offer several statements. Polyculturalism is
formed by different ways of thinking that can be split into two
components: the intuitive-creative or imaginative thinking and the
heuristic-logical thinking which can be called intellectual thinking.
Further, for describing the intellectual-creative activity we shall make use
of the notion of the state amplitude. It is a value which when squared
gives  the probability of staying in this state. Therefore, we tend to believe that
there are only two reference states which can be called the state of
imaginative thinking and the state of logical thinking.  Each of these
states is described by its own amplitude. Likewise, polyculturalism can
also be described as an overlap in varied proportions of just two
constituents. By this we suppose that polyculturalism is two-dimensional.
This resembles a well-known fact that any colour can be split into three
constituents: red, green and blue. Similarly, mathematicians believe that
any vector can be split into three reference vectors. This is tantamount  to
saying that in our case each vector in the polycultural space
determines the state vector which in its turn is subject to
decomposition into two reference vectors.

\begin{center}
IV
\end{center}

The aforementioned preliminary remarks, at times, if anything, not
very clearly defined, will now be formulated as several postulates
developing our quantum mechanics approach and at the same time
introducing the necessary notions and respective mathematical
operations.

\vspace*{1.0cm}

\noindent
{\bf Postulate I}. {\em Let the state of the system's intellectual-creative activity at the
microscopic level be given by the amplitude $\psi$ dependent upon the time
$t$ and ``the generalized coordinates"  $q$. The module square of the value $\psi$
equals the probability of the system's staying in this state.}

\vspace*{1.0cm}

\noindent
{\bf Postulate II}. {\em The space of the intellectual-creative activity is
believed to be two-dimensional: one dimension is imaginative
thinking, the other one is logical thinking. Let $\psi_1$ be
the amplitude  of the state of imaginative thinking and let $\psi_2$ be the amplitude
of the state of logical thinking, the  $\psi_1$, $\psi_2$ base being
complete.}

\vspace*{1.0cm}

\noindent
{\bf Postulate III}. {\em The amplitude of any state is determined in
compliance with the quantum mechanics principle of superposition:
\ben
\psi=C_1\psi_1+C_2\psi_2.
\een
%(1)
Between the states $\psi_1$ and  $\psi_2$  describing  the two
cultures transitions are possible which are characterized by the
so-called exchange energy A}.

The dependence of the amplitude $\psi$ upon time is determined by
the factors  $C_1$  and  $C_2$ which can be found from Schr\"odinger's equation
that is the main equation of quantum mechanics [5,6]:
\ben
C_1&=&\cos\Omega t,\nonumber\\
C_2&=&-i\sin\Omega t,
\een
where the frequency of ``exchange" between the reference states $\Omega=A/\hbar$ is a
characteristic parameter of the system and is determined both by the
intrinsic properties  of the system itself and its environment. For the sake
of definiteness it is taken that at the initial moment of time $t=0$  the
system is in the state $\psi_1$ . Besides this, the given expressions do not take
into account the time dependent phase multiplier which is irrelevant for
us and which is removed by a simple replacement of the beginning of the
measurement on the energy scale.

The full probability of staying in the two-cultural state equals the
module square of the amplitude $\psi$. With this full probability we shall
calculate the aggregate product of the intellectual-creative capacity of a
person
which stands for what one manages to create in the
course of the entire creative life.

Let the intellectual-intuitive creative capacity be described by the
operator
\ben
\hat K=\hat K(q,t)
\een
in the way that its observable value at the moment $t$, {\it i.e.}, the quantum
mechanical average of the said operator by definition  equals
\ben K(t)=\la \psi|\hat K|\psi\ra=\int\psi^{*}(q,t)\hat K(q,t)\,\psi(q,t)\,dq.
\een

 In the course of the creative life time span $T$ of a person the
aggregate product equals  the
integral  of the average capacity
\ben Q=\int\limits_0^T K(t)\,dt.\een

The beginning $t=0$ when the creative activity gets off the ground is
dependent upon the IQ coefficient [3]. The end of the creative life when
$T=t$ and
when one exhausts oneself is dependent on many factors, not the least
quite accidental.

 With the purpose of simplifying the notation without losing the
final conclusions we shall separate in the operator $\hat K$ the dependence
upon time $t$ and the coordinates $q$:
\ben
\hat K(q,t)=\hat Q(q)p(t),
\een
at the same time we shall normalize the function $p(t)$ so that
\ben
\int\limits_0^T p(t)\, dt=1.
\een

We will place this operator in the expression for $K(t)$ and taking into
account the explicit form of the amplitude $\psi$ and the coefficients $C_1$ and
$C_2$ we will find that
\ben
K(t)&=&p(t)\left[Q_{11}\cos^2 \Omega t+Q_{22}\sin^2\Omega t
+i(Q_{21}-Q_{12})\sin \Omega t\cos \Omega t\right],
\een
where the  matrix elements of the operator $\hat K$ equal:
\ben
Q_{11}&=&\la\psi_1 |\hat Q|\,\psi_1\ra,\nonumber\\
Q_{22}&=&\la\psi_2 |\hat Q|\,\psi_2\ra,\\
Q_{12}&=&Q_{21}^{*}=\la\psi_1 |\hat Q|\,\psi_2\ra.\nonumber
\een

By its content the operator $\hat K$ is self-correlative and expressed positively,
thus $Q_{11}>0$, $Q_{22}>0$, the non-diagonal element being  written as a
complex value through its module and $\delta$ phase.
\ben
Q_{12}=|Q_{12}|e^{i\delta},
\een
note that since  $Q_{21}=Q_{12}^*$,
\ben
Q_{21}=|Q_{12}|e^{-i\delta}.
\een
Now,  the formula for $K(t)$ can be written as follows
\ben
K(t)=p(t)\left({Q_{11}+Q_{22}\over 2}+{Q_{11}-Q_{22}\over 2}\cos 2\Omega t
+|Q_{12}|\sin\delta\, \sin 2\Omega t\right).
\een
As we can see an essential dependence upon the phase $\delta$
arises. Integrating this expression by the time $t$ we have
\ben
Q={Q_{11}+Q_{22}\over 2}+{Q_{11}-Q_{22}\over
2}p_{2\Omega}'+|Q_{12}|p_{2\Omega}''\sin \delta,
\een
where we have introduced the real and imaginary parts of the Fourier
coefficient of the function $p(t)$:
\ben
&&p_{\omega}=\int\limits_0^T p(t)e^{i\omega t},\\
&&p_{\omega}'={\rm Re}\,p_{\omega},\ \ \ \ p_{\omega}''={\rm
Im}\,p_{\omega}.\nonumber
\een
The found expression for $Q$ happens to be the main formula
providing the basis for the entire subsequent analysis.

\begin{center}
{\bf V}
\end{center}

Rather than seeing our task in comparing the intellectual capacities of
different people  we are primarily concerned with finding out more about
the circumstances under which one can fully implement one's own abilities
in compliance with the individual's own potential. That is why it is natural
to assume the measuring unit of the value $Q$ to be  a half of the sum
$(Q_{11}+Q_{22})/2$ which we will refer to as a  classical meaning of the aggregate
 product. Thus, we will be concerned with the scaled aggregate
product
\ben
Q^{*}=Q\Bigg/\left({Q_{11}+Q_{22}\over 2}\right). \een
Proceeding from the main formula we will receive
\ben
Q^{*}=1+{Q_{11}-Q_{22}\over Q_{11}+Q_{22}}p_{2\Omega}'+{2|Q_{12}|\over Q_{11}+Q_{22}}p_{2\Omega}''\sin\delta.
\een
As we treat of a person equally endowed with talents, {\it i.e.},
when $Q_{11}=Q_{22}$
this expression is simplified:
\ben
Q^{*}=1+{|Q_{12}|\over Q_{11}} p_{2\Omega}''\sin \delta.
\een

Let us pass over to the discussion of the established formula. The
maximum possible exhaustedness of a creative personality equals a half
of the sum of what originates from each of the two reference states
(otherwise said from the intellectual and creative states)  together with
the interference crossed term. The crossed term which is refuted by the
classical logic appears, as can be seen, from the doubled product  during
the unfolding of the square from $\psi$ as a sum of two terms at the calculus
of probability. This interferential contribution into the aggregate product
is dependent on two parameters: the frequency $\Omega$ of switching over from
one type of activity  to the other as well as on such a subtle, almost
transcedental notion as the phase $\delta$.

The crossed effect can enhance or decrease the  implementation of the
person's potential depending upon these parameters.
In order to implement one's talent to the full it is necessary first and
foremost to harmonize the switching over frequency with the length of
the creative life so that the value of $p_{2\Omega}''$  be maximally positive. Secondly,
we must set the phase $\delta$ to equal $\pi/2$, so $\sin\delta=1$. Then the contribution
of two-culturalism into the aggregate product will  be not just
maximal  but will also have a positive sign.

If the phase $\delta$ equals zero or the switching over frequency
is high   $\Omega\to \infty$, $p_{2\Omega}''\to 0$, a gifted person does not tend to fully unfold
one's own abilities and turns into an ordinary person with half a sum of  the
aggregate product $(Q_{11}+Q_{22})/2$ realizing  the classical values of $Q^{*}=1$.
Should the phase  $\delta$ equal $3\pi/2$, we will have a sign which is opposite to
two-culturalism and then the person in question is not merely a lost
talent. We may have to deal with a tragedy as a gifted person has
failed to come up to the achievements of an ordinary person.

Hence, purposeful efforts are needed to obtain self-implementation
and ensure harmony with oneself as well as intellectual and aesthetic
satisfaction. These efforts should  be directed at setting the input parameters
characterizing the intellectual creative activity of a person.

\begin{center}
{\bf VI}
\end{center}

 With the purpose of giving
numerical characteristics of these qualitative conclusions we will study the model systems. Let us first
consider the symmetrical case when $|Q_{12}|=Q_{11}=Q_{22}$. We will give
$p(t)$ as a step function, {\it i.e.}, let us assume that $p(t)$ remains constant from the
beginning of one's creative life $t=0$ till $T=0$ when the creative life comes
to an end:
\ben
p(t)=\left\{
\begin{array}{ll}
{1/T}, & 0\le t\le T\\
0, &  otherwise,
\end{array}
\right.,
\een
the Fourier coefficient
\ben
p_{\omega}={1\over T} \int\limits_0^T e^{i\omega t}\,
dt={e^{i\omega T}-1\over i\omega T},
\een
and the imaginary part
\ben
p_{\omega}''={\sin^2(\omega T/2)\over (\omega T/2)}.
\een
Consequently,
\ben
Q^{*}=1+\sin \delta {\sin^2 \Omega T\over \Omega T}.
\een
The maximum value of $Q^{*}$  brings about $\delta=\pi/2$ as well as the parameter
$x=\Omega T$ which complies by the equation  (the condition according to which
the derivative from $Q^{*}$ by $x$ equals zero)
\ben
{{\rm tg}\, x\over x}=2,
\een
from where $x=1.165561$. At this
\ben
Q_{\rm max}^{*}=1+{\sin^2 x\over x}=1.7246.
\een

Hence, the interference member tends to additionally increase the classical value of
$Q^{*}$ by 72\%. It is also curious that in this model when one's creative life is cut
short at once at $x=n\pi$, $n=1,2,\ldots$ interference disappears and the value of
$Q^{*}$ becomes classical too. If the frequency of switching over $\Omega\to\infty$,
then $Q^{*}\to 1$. For that matter  a frequent change of interests tends to smear the
interference effect of one's intellectual-creative activity, the said effect
being the essence of one's unusualness.
When a person is able to set the parameters of $\delta$ and $\Omega$ to their
extreme values he possesses a many-sided and well manifesting itself
talent.

Let us now consider a more realistic model of the function $p(t)$. Let
$p(t)$ be the analytical function of the time pending to zero both  at $t\to 0$ and
at $t\to\infty$:
$$p(t)=\gamma^2 t e^{-\gamma t},$$
where $\gamma$ is a  decrement  of the subsidence of the intellectual-intuitive creativity whose
presence allows to detract  the parameter $T$ to infinity  ($T=\infty$). In
compliance with our requirements  this function is  normalized by one,
the climax of the creative activity falling on the time $t=t/\gamma$.
    By definition we shall calculate the Fourier coefficient
\ben
p_\omega={1-(\omega/\gamma)^2+2i\omega/\gamma\over
\left[1+(\omega/\gamma)^2\right]^2}.
\een
Let us take from here $p_{\omega}''$ at the frequency $\omega=2\Omega$
and for $Q^{*}$ we will have:
\ben
Q^{*}=1+{2\nu\over (1+\nu^2)^2}\sin \delta,
\een
where $\nu=2\Omega/\gamma$. This function also falls  to zero at $\nu\to\infty$. And it reaches its peak
at $\delta=\pi/2$ at the point $\nu=1/\sqrt 3$:
\ben
Q_{\rm max}^{*}=1+3\sqrt 3/8=1.6495.
\een
The interference effect in this model increases the classical value
of $Q^{*}$ by 65$\%$.

\begin{center}
{\bf VII}
\end{center}

It seems likely that each of us carries his own {\it problem of two
cultures} inside himself. Both an artist and a scholar often act
without the awareness that they themselves are at the crossroads of these
cultures. The suggested quantum-mechanical mechanism of creativity
gives a paradoxical intertwining of imaginative and logical thinking.
This intertwining is capable of creating something which looks entirely
new and cannot be reduced to anything or split into constituents. It
should rather be seen as an interference effect. The contribution of
two-culturalism into the aggregate product understood as the
interferencial crossed effect can alter slightly depending upon the
specific model. Yet it is this effect which makes the difference between
potentially talented people and people that implement their talent.

Ultimately, let us remark that these conclusions can pertain not
only to specific individuals who find themselves in two states at a time
but also to a community of people that are forced by the historical
circumstances to live and act in the conditions of multiple cultural
cross-overs.

One  of the two  quotations  placed at the beginning of this paper is
from the work of the outstanding Ukrainian thinker Hryhorij Skovoroda.
It refers us once again to the fundamental philosophical principle of
getting to know oneself which has always been attractive to people even
though with varied intensity at different times. With an ever-growing vigour
we are obliged to get to know our own selves,  researching and opening
anew our activities in versatile spheres. It is not only quite interesting in
itself but also increasingly important in view of the fact that we happen
to live in an ever more interrelated and globalized society when the fate
of each of us is dependent upon the activity of everyone else.

\end{document}